\documentclass[showpacs,amsmath,amssymb,superscriptaddress]{revtex4}
\usepackage{graphicx}
\usepackage{dcolumn}
\usepackage{bm}

\begin{document}
\title{Total suppression of a large  spin tunneling barrier in
quantum adiabatic computation}
\author{A. Boulatov}
\email{boulatov@email.arc.nasa.gov}
 \affiliation {QSS Group, Inc;  NASA Ames Research
Center, MS 269-2, Moffet Field, CA 94035-1000}
\author{V.N. Smelyanskiy}
\email{vadim@email.arc.nasa.gov}
 \affiliation
{NASA Ames Research
Center, MS 269-2, Moffet Field, CA 94035-1000}

\date{\today}
\begin{abstract}
  We apply a quantum adiabatic evolution algorithm to a combinatorial
  optimization problem where the cost function depends entirely on the
  of the number of unit bits in a $n$-bit string (Hamming weight).  The
  solution of the optimization problem  is encoded as a
  ground state of the problem Hamiltonian $H_{p}$
  for the $z$-projection of a total spin-$\frac{n}{2}$.  We show that
  tunneling barriers for the total spin can be completely suppressed
  during the algorithm if the initial Hamiltonian has its ground state
  extended in the space of the $z$-projections  of the spin.
  This suppression takes place even if the cost function has deep and
  well separated local minima.  We provide an intuitive picture for
  this effect and show that it guarantees the polynomial complexity of
  the algorithm in a very broad class of cost functions. 
  We suggest a simple example of the Hamiltonian for
  the adiabatic evolution: $H(\tau)=(1-\tau)\, \hat S_{x}^{2}
  +\tau\, H_{p}$, with parameter $\tau$ slowly varying in time
  between $0$ and $1$.  We use WKB analysis for the large spin to
  estimate the minimum energy gap between the two lowest adiabatic
  eigenvalues of $H(\tau)$.
\end{abstract}

\maketitle

\section{\label{sec:Intro}Introduction}

Recently a novel paradigm was suggested for the design of quantum
algorithms for solving combinatorial search and optimization problems
based on quantum adiabatic evolution \cite{Farhi}.  In the quantum
adiabatic evolution algorithm (QAA) a quantum state  closely
follows a ground state of a specially designed slowly varying in time
control Hamiltonian.  At the initial moment of time the control
Hamiltonian has a simple form with the known ground state that is easy
to prepare; at the final moment of time it coincides with the \lq\lq problem''
Hamiltonian $H_P$ whose  ground state encodes the
solution of the classical optimization problem in question.
It can also  be chosen to reflect the
bit-structure and cost spectrum of the problem. For example,
\begin{eqnarray}
&& H_P=\sum_{\bf z} E_{\bf z}  | {\bf z}\rangle   \langle {\bf z}|
\label{HP} \\ && |{\bf z}\rangle = | z_1\rangle_1\,
\otimes|z_2\rangle_2\,\otimes\cdots\otimes|z_n\rangle_n .
\nonumber
\end{eqnarray}
\noindent Here $E_{\bf z}$ is a cost function defined on a set of
$2^n$ binary strings ${\bf z}=\{z_1,\ldots, z_n \}$ $z_j=0,1$, each
containing $n$ bits. The summation in (\ref{HP}) is over $2^n$ states
$|{\bf z}\rangle$ forming the computational basis of a quantum
computer with $n$ qubits. State $|z_j\rangle_j$ of the $j$-th qubit is
an eigenstate of the Pauli matrix $\hat{\sigma}_z$ with eigenvalue
$1-2z_j= \pm 1)$.  If at the end of the QAA the quantum state is
sufficiently close to the ground state of $H_P$ then the solution to
the optimization problem can be retrieved by  measurement.

Running of the algorithm for several NP-complete problems has been
simulated on a classical computer using a large number of randomly
generated problem instances that are believed to be computationally
hard for classical algorithms \cite{FarhiSc,FarhiSat,FarhiCli}.
Results of these numerical simulations for relatively small size of
the problem instances ( $n \leq$ 20) suggest a {\em quadratic} scaling
law of the run time of the quantum adiabatic algorithm with $n$.
Furthermore, it was shown in \cite{Vazirani02} that the previous query
complexity argument that lead to the exponential lower bound for 
unstructured search \cite{Bennett} cannot be used to rule out the
polynomial time solution of NP-complete Satisfiability problem by
QAA.

On the other hand, a set of examples of the 3-Satisfiability problem
has been recently constructed \cite{Vazirani01,annealing,Vazirani02}
to test analytically the power of QAA in the situations where the
optimization problem in question has multiple well-separated local
minima.

In these examples the cost function $E_{\bf z}$ depends on a
bit-string ${\bf z}$ with $n$ bits, ${\bf z}=\{z_1,z_2,\ldots,z_n\}$,
only via the Hamming weight of the string, $w_{\bf
  z}=(z_1+z_2+\ldots+z_n)$, so that $E_{\bf z}=f\left(w_{\bf z}\right)$. The
function $f(w)$ is multi-modal, it has a local minimum separated
from the global minimum by the barrier of an order-$n$ width in
$w$. For that reason classical local search like simulating
annealing provably fails to find a globally optimal solution in
time polynomial in $n$. It \cite{annealing,Vazirani02} a \lq\lq
standard'' QAA was applied to this problem in which a quantum
evolution begins in a uniform superposition state
$\frac{1}{\sqrt{2^n}}\sum_{\bf z}|{\bf z}\rangle$, ends in a
target (solution) state $|{\bf z}_{t}\rangle$, and the control
Hamiltonian is a linear interpolation in time between the initial
and final Hamiltonians.

For the above examples it was shown \cite{annealing,Vazirani02}
that the  system can be trapped during the QAA in the local
minimum of a cost function for a time that grows exponentially in
the problem size $n$.  It was also shown \cite{annealing} that an
exponential delay time in QAA can be
computed in terms of a quantum-mechanical tunneling for an
auxiliary large spin system.

It can also be inferred from \cite{Vazirani02,annealing} that QAA
will have an exponential complexity even if the cost function
$E_{\bf z}$  no longer depends strictly on a Hamming weight but
the deviation only occurs for states $|{\bf z}\rangle$ that
have exponentially small (in $n$) overlap with the adiabatic
ground state wavefunction $|\phi_{0}(t)\rangle$ at all times
during the algorithm execution.

The above example has a significance more than just being a
particular simplified case of the binary optimization problem with
symmetized cost. Indeed one can argue that it shows one of the
mechanisms for setting \lq\lq locality traps'' in the
3-Satisfiability problem \cite{Vazirani_talk}.  But most
importantly, this example demonstrates that exponential complexity
of QAA results from a {\em collective
  phenomenon} in which transitions between the bit-configurations with
low-lying energies can only occur by the simultaneous flipping of
large clusters containing order-n bits.  In many cases these
transitions can be analyzed as a tunneling of spin variables. A
similar phenomenon related to the tunneling of magnetization was
recently observed in the large-spin molecular nanomagnets
\cite{Wernsdorfer}.

However low-energy collective behavior is also well known in spin
glass models, many of which are in one-to-one correspondence with
random NP-complete problems \cite{Anderson}. In particular, an
important ingredient of the \lq\lq replica symmetry breaking''
picture of an infinite-range spin glass by Parizi \cite{Parizi} is
that there are collective spin exitations that are of the order of
the system size $n$ whose energy is $({\cal O}(1)$, i.e., it
does not grow with the size of
the system. A similar picture may be applicable to  random
Satisfiability problems \cite{Monasson}.

Therefore in connection to the above example, it is important to
understand how to design a polynomial time QAA without a prior
knowledge of a particular form of the cost function
$f\left(w\right)$ (possibly multi-modal or even randomly sampled),
so that a tunneling barrier between the local minima will be
totally suppressed. This is a focus of the present paper
\cite{footnote1}.

In Sec.~\ref{sec:spin} we present a theory of  quantum
adiabatic evolution of a large spin system introducing a control
Hamiltonian that guarantees  polynomial time complexity for QAA
in the symmetrized 3-Satisfiability example mentioned.  In
Sec.~\ref{sec:gap1} we provide an estimate of the minimum gap
between the two lowest eigenvalues of the control Hamiltonian that
determines the complexity of QAA.  Sec.~\ref{sec:conclusion}
contains concluding remarks.

\section{\label{sec: QAA} Quantum Adiabatic Evolution Algorithm}

 In a standard  QAA   \cite{Farhi} one specifies
the time-dependent Hamiltonian $H(t)=\tilde H(t/T)$
\begin{equation}
\tilde H(\tau) = (1-\tau)\, H_D+ \tau\, H_P,  \label{H_tot1}
\end{equation}
\noindent where $\tau=t/T$ is dimensionless \lq\lq time". This
Hamiltonian guides the quantum evolution of the state vector
$|\psi(t)\rangle$ according to the Schr{\" o}dinger equation $
i\,{\partial |\psi(t)\rangle \partial t} =H(t) |\psi(t)\rangle$
from $t=0$ to $t=T$, the {\em run time} of the algorithm (we let
$\hbar=1$). $H_P$ is the \lq\lq problem" Hamiltonian given in
(\ref{HP}). $H_D$ is a \lq\lq driver" Hamiltonian, that is
designed to cause the transitions between the eigenstates of
$H_P$. In this algorithm one prepares the initial state of the
system $|\psi(0)\rangle$ to be a ground state of $\tilde
H(0)=H_D$. It is typically constructed assuming ${\it no}$
knowledge of the solution of the classical optimization problem
and related ground state of $H_P$. In the simplest case
\begin{equation}
 H_D = -C\,\sum_{j=0}^{n-1} \hat \sigma_{x}^{j},\quad
|\psi(0)\rangle =2^{-n/2}\sum_{\bf z}|{\bf z}\rangle, \label{HD}
\end{equation}
\noindent  where $\sigma_{x}^{j}$ is a Pauli matrix for $j$-th
qubit and $C>0$ is some scaling constant.  Consider instantaneous
eigenstates $|\phi_{k}(\tau)\rangle$ of $\tilde H(\tau)$ with
energies $E_{k}(\tau)$ arranged in nondecreasing order at any
value of $\tau\in(0,1)$
\begin{equation}
\tilde  H |\phi_{k}\rangle = E_{k} |\phi_{k}\rangle, \quad
k=0,1,\ldots,2^{n}-1.\label{adiab}
\end{equation}\noindent
 Provided the value of $T$ is large enough and there is a finite gap
for all $t\in(0,T)$  between the
 ground and exited state energies,
 $\Delta E(\tau)=E_1(\tau)-E_0(\tau)>0$,
 quantum evolution is adiabatic and the state of the system
 $|\psi(t)\rangle$ stays
close to an instantaneous ground state, $|\phi_0(t/T)\rangle$ (up
to a phase factor). Because  $H(T)=H_P$ the final state
$|\psi(T)\rangle$ is close to the ground state
$|\phi_0(\tau=1)\rangle$ of the problem Hamiltonian. Therefore a
measurement performed on the quantum computer at $t=T\, (\tau=1)$
will find one of the solutions of combinatorial optimization
problem  with large probability. Quantum transition away from the
adiabatic ground state occurs most likely in the vicinity of the
point $\tau\approx \tau_c$ where the energy gap $\Delta E(\tau)$
reaches its minimum (avoided-crossing region). The probability of
the transition, $1-|\langle
\psi(t)|\phi_0(t/T)\rangle|^{2}_{t=T}$, is small provided that
\begin{equation}
T\gg  \frac{ |\langle \phi_{1}|\tilde
H_{\tau}|\phi_{0}\rangle|_{\tau=\tau_c} }{\Delta E_{\rm min}^{2}}
,\quad \Delta E_{\rm min}=\min_{0\leq \tau\leq
1}\left[E_1(\tau)-E_0(\tau)\right], \label{mingap}
\end{equation}
\noindent ($\tilde{H}_{\tau} \equiv d\tilde{H}/d\tau$). The
fraction in (\ref{mingap}) gives an estimate for the required
runtime of the algorithm and the task is to find its asymptotic
behavior in the limit of large $n \gg 1$.  The numerator in
(\ref{mingap}) is less than the largest eigenvalue of
$\tilde{H}_{\tau}=H_P - H_D$, typically polynomial in $n$
\cite{Farhi}. However, $\Delta E_{\rm min}$ can scale down
exponentially with $n$ and in such cases the runtime of quantum
adiabatic algorithm will grow exponentially fast with the size of
the input $n$.

\subsection{\label{sec:symm}Binary optimization problems with
symmetrized cost function}
Consider a cost function $E_{\bf z}$ in the following form:
\begin{equation}
E_{\bf z}=f\left(w_{\bf z}\right), \quad w_{\bf z}=\sum_{j=1}^{n} z_j.
\label{symcost}
\end{equation}
\noindent
This cost is symmetric with respect to the permutation of bits and
$w_{\bf z}$ is a Hamming weight of a string ${\bf z}$.  A particular
example of this problem related to 3-Satisfiability was introduced in
\cite{Vazirani01,annealing,Vazirani02} (the discussion in this subsection
closely follows \cite{annealing}, Sec.~4).  In this example
\begin{eqnarray}
&& E_{\bf z}=\sum_{i<j<k}c(z_i+z_j+z_k), \nonumber \\
&& c(m)=(1-\delta_{m,0})\,\left (1+\delta_{m,1}(q-1)\right), \quad m=0,1,2,3\label{3sat}
\end{eqnarray}
\noindent where $\delta_{k,l}$ is a Kronecker delta. For this
particular case function $f(w)$ in (\ref{symcost})
 takes the following form:
\begin{equation}
f(w)=\frac{q}{2}w(n-w)(n-w-1)+\frac{1}{2}w(w-1)(n-w)+\frac{1}{6}w(w-1)(w-2).\label{f0}
\end{equation}
\noindent
where $q$ is an integer greater than or equal to 3.
In the leading order in $n\gg 1$ one can write:
\begin{equation}
f(w)=\left(\frac{n}{3}\right)^3\, g\left(\frac{w}{n}\right)+{\cal O}(n^2),
\label{f1}
\end{equation}
\noindent
where
\begin{equation}
h(u)=4qu(1-u)^{2}+4u^{2}(1-u)+\frac{4}{3}u^{3}.\label{h}
\end{equation}
\noindent $g(u)$ is a  non-monotonic function with  global minimum
at $u=0$ corresponding to $z_1,z_2,\ldots,z_n=0$. It also has a
local minimum at $u=1$ corresponding to $z_1,z_2,\ldots,z_n=1$ (
cf. Ref.~\cite{annealing}, Fig.~1).

In QAA the symmetrized cost function (\ref{f0}) corresponds to the
following problem Hamiltonian $H_P$ of  total spin-$n/2$ system
(cf. Eq.~(\ref{HP}) )
\begin{equation}
H_P=f\left(\frac{n}{2}-\hat S_{z}\right),\label{3satHP}
\end{equation}
\noindent
where
\begin{equation}
\hat S_{z}|{\bf z}\rangle = S_z )|{\bf z}\rangle, \quad S_z=
\frac{n}{2}-w_{\bf z}.\label{Sz}
\end{equation}
\noindent Here $\hat S_z$ is operator of $z$-projection of a total
spin $\frac{n}{2}$  of a system of $n$ individual spins $1/2$. We
used an obvious connection between the values of the Hamming
weight function $w_{\bf z}$ and corresponding eigenvalues $S_z$ of
the operator $\hat S_z$. In what following we use \lq\lq hat'' notation
for the  operators of  a total spin.

It was shown \cite{annealing} that if one uses $H_D$ in a form
corresponding to (\ref{HD}) 
\begin{equation}
H_{D}=\binom{n-1}{2}\left(\frac{n}{2}-\hat S_{x}\right).\label{HDfarhi}
\end{equation}
\noindent then the minimum gap $\Delta E_{\rm min}$ scales down
exponentially with $n$ implying the exponential complexity of QAA
for this problem.

\section{\label{sec:spin} Adiabatic Evolution of a Large Spin}

\subsection{Extended {\it vs} localized initial states}
 In this paper we
show that the main reason for the exponentially small minimum gap
 for the problem (\ref{3sat}) with $H_D$ given in (\ref{HDfarhi})
 is the fact that the ground state of
the driving Hamiltonian is {\bf localized} in the space of the
$z$-projections of a total spin.  We construct an example of the
driving Hamiltonian with {\bf extended} ground state and show that
in this case the evolution time of QAA is polynomial in the number
of qubits.

In the following we adopt the notation for the total spin $\hat
{\bf S}= \left\{ \hat {S}_{j}\right\} $, with $j=x,y,z$, where
$\hat S_{j}$ are the projections of the total spin operator on the
$j$-th axis. The total spin operator's components equal $\hat
S_{j}=\sum_{k=1}^{n}\hat S_{j}^{\left( k\right) }$ and are
symmetric in all one-qubit spin operators $\left\{ \hat
S_{j}^{\left( k\right) }\right\}$ for $k=1,...n$.
 We use
$S_j$ for the eigenvalues of the operator $\hat S_j$.

We now introduce a new driver Hamiltonian
\begin{equation}
H_{D}=\hat S_{x}^{2}.\label{driverSx2}
\end{equation}
\noindent Consider, for example, its ground state in the case when the
total number of qubits $n=2s$ is even and therefore the total spin
$l=n/2=s$ is an integer. Then the ground state is a state with
$S_{x}=0$ in a basis where $x$ is chosen as a quantization axis.
Making use of  Wigner's rotation matrix $d_{m,m^{\prime
}}^{l}\left( \theta \right) $ \cite{LANDAU1}, one easily projects
this state onto the computational (problem) basis with the $z$
quantization axis and $m=S_z$. This gives us the ground state wave
function in the problem basis as $\Psi _{0}\left( m\right)
=d_{m,0}^{l}\left( \pi /2\right) $. In the limit of large spin
(which is the case of interest for us), the wave function is given
by

\begin{equation}
\Psi _{0}\left(m\right) \approx \frac{\left( -1\right) ^{k}}{\sqrt{2\pi }}%
\frac{1}{\left( l^{2}-m^{2}\right) ^{1/4}}\delta _{l-m,2k}, \qquad
l=\frac{n}{2} \quad (n\,\,{\rm  is \,\,even}). \label{state_1}
\end{equation}
\noindent Note that $\Psi _{0}\left( m\right) =0$ for the odd
difference $l-m=2k+1$ and is given by (\ref{state_1}) for
$l-m=2k$. From (\ref{state_1}), it follows that the ground state
of $S_{x}^{2}$ is delocalized in the $m$-
 space and spread over the whole interval, $-l\leq m\leq
l$.

On the other hand, the ground state of the operator
(\ref{HDfarhi}) used  in \cite{annealing} as a driver, is
localized in the $m$-space. Indeed, that ground state is a state
with $S_{x}=l$ in the $x$ basis. This gives us the ground state
wave function in the problem basis,  $\Psi^{\prime}_{0}\left(
m\right) =d_{m,l}^{l}\left( \pi /2\right) $. In the limit of large
spin one has
\begin{equation}
\Psi^{\prime} _{0}\left( m\right) \approx \frac{1}{\sqrt{\pi l}}\exp \left( -\frac{
m^{2}}{l}\right) ,  \label{state_2}
\end{equation}
\noindent which is clearly localized on the scale $m^{2}\simeq l$.
The same conclusion holds for the case when $n$ is odd and $l$
take half-integer values.

As we will show below, the adiabatic evolution of the delocalized
states is fundamentally different from the localized ones. In
particular, the localized ground states in general result in 
macroscopic tunneling. This gives the exponentially small in $n$
ground state energy gaps and consequently the exponentially large
complexity of QAA. Using the delocalized states one can avoid 
macroscopic tunneling. We argue that in a general situation when
the information about  the ground state is not used for constructing
the driver Hamiltonian, the driver with extended ground state
should lead to polynomial complexity of adiabatic algorithms
independently of the specific form of the problem Hamiltonian
provided that it is expressed as a function (in general,
nonlinear) of the total spin operators, $\hat S_{j}$ (cf.
Eq.~(\ref{3satHP})).

\subsection{WKB approximation for the large spin}

To be specific, we refer to the same problem Hamiltonian as in
Eq.~(\ref{3satHP}). The full Hamiltonian takes the following form
in the limit of large $n$
\begin{eqnarray}
&&H\left( \tau \right) =\left( 1-\tau \right) n \hat S_{x}^{2}+\tau \left( \frac{n}{2%
}\right) ^{3}h\left( \hat u\right) ,  \label{S_x_21}\\ &&
\hat u=\frac{1}{2}\left( 1-\frac{\hat S_z}{l}\right), \qquad
l=\frac{n}{2},\nonumber
\end{eqnarray}
where $\tau =t/T$ ; $0\leq \tau \leq 1$ and function $h(u)$ is
given in (\ref{h}).

 In order to get a simple physical picture, we
will refer to the WKB-type approach commonly used in the theory of
quantum spin tunneling in magnetics (QTM) \cite{CHUD1},
\cite{GARG1}, \cite{GARG2}. This approach is applicable for 
large spins $(l>>1)$, which is the case of interest for us. We
choose $x$ as a quantization axis and following the standard
procedure obtain the effective quasi-classical Hamiltonian in
polar coordinates $\left\{ \theta ,\varphi \right\} $ with $\theta
\in \left[ 0,\pi \right] $ and $\varphi \in \left[ 0,2\pi \right]
$. In doing this, we make use of the following relations valid in
the limit $(l>>1)$
\[
S_{z}^{n}\approx \left( l\sin \theta \sin \varphi \right) ^{n}.
\]
\noindent
As it was shown in \cite{GARG2}, in the quasi-classical limit we have 
\begin{equation}
\hat S_{x}^{2}\approx - \frac{ \partial^2}{\partial \varphi^2} \equiv \hat p^{2},
\quad  \theta \approx \frac{\pi}{2},
\end{equation}
\noindent
which means that the motion of the
quasi-classical spin can be described as a 1D motion of a massive
particle on a unit radius ring in the appropriate effective
potential $V(\phi)$. We also have
\begin{equation}
u=\frac{1}{2}\left( 1-\sin \varphi \right) .  \label{u_1}
\end{equation}
\noindent
Substituting this into (\ref{S_x_21}), we finally obtain the
effective quasi-classical Hamiltonian in the form
\begin{eqnarray}
&&H\left( \tau \right) =n\left( 1-\tau \right) \hat p^{2}+\tau \left( \frac{n}{2}%
\right) ^{3}\, V(\varphi), \nonumber \\&& \quad V(\varphi)=h\left(
\frac{1-\sin \varphi }{2}\right) . \label{H_qc_1}
\end{eqnarray}
\noindent
One should note that the driver $H_{D}=n \hat S_{x}^{2}$ does not change the
effective potential caused by $H_{P}$ and only introduces the effective
kinetic energy into the problem. As we show below, this driver has extended
eigenstates in the space of the $z$-projections $m$ of a total spin, 
as opposed to the case $H_{D}=n^{2}S_{x}$
considered in \cite{annealing}, where the eigenstates are localized in $m$. As
we will see below, this leads to the absence of tunneling and polynomial gap
in our case as opposed to the exponentially small  tunneling amplitude
arising in \cite{annealing}
. The adiabatic wave functions $\Psi _{k}\left( \varphi \right) $ satisfy
\begin{equation}
\left[ -\left( 1-\tau \right) \frac{\partial }{\partial \varphi
^{2}}+\tau \frac{n^{2}}{8}\,V(\varphi)  \right] \Psi _{k}\left(
\varphi \right) =\widetilde{E}_{k}\left( \tau \right) \Psi
_{k}\left( \varphi \right) ,  \label{Schr_qc_1}
\end{equation}
\noindent
where $\widetilde{E}_{k}\left( \tau \right) =E_{k}\left( \tau \right) /n$ is
a rescaled energy. The first term in the l.h.s. of (\ref{Schr_qc_1})
corresponds to the driving Hamiltonian $H_{D}$ and presents the kinetic
energy of the particle moving in the effective potential due to the problem
Hamiltonian $H_{P}.$  At initial moment $\tau =0$, the total Hamiltonian reduces to
the driver $H_{D}$ which describes a free massive particle on a ring and its
eigenstates are well known \cite{LANDAU1}. In this case, the Schr\"{o}dinger
equation (\ref{Schr_qc_1}) gives exact wavefunctions and spectrum. Namely,
we have

\begin{eqnarray}
\Psi _{k}\left( \varphi \right) &=&\frac{1}{\sqrt{2\pi }}\exp \left(
ik\varphi \right) ,  \label{Schr_qc_2} \\
\widetilde{E}_{k}\left( 0\right) &=&k^{2},  \nonumber
\end{eqnarray}
\noindent
where $k$ is an integer number $k=0,\pm 1,\pm 2,...$ for even $n$ when
the total spin $l=n/2$ is integer, and a half-integer number $k=\pm 1/2,\pm
3/2,...$ for an odd number of qubits $n$, when $l=[n/2]+1/2$. One should
note that in case of integer spin $l$ , all eigenstates are twofold
degenerate except for the ground state corresponding to $k=0$ and for the
half-integer spin, all states including the ground state are twofold
degenerate. This is a particular case of the Kramers' degeneracy \cite
{LANDAU1}, \cite{GARG1}, which occurs due to the symmetry with respect to
the spin-flip transformation. Note that since from (\ref{u_1}) it follows
that $m=l\sin \varphi $, the states (\ref{Schr_qc_2}) are indeed delocalized
in the $m$-space.

From (\ref{Schr_qc_1}), it follows that the problem Hamiltonian is of the
same order of magnitude as the driving one only for sufficiently short times
$\tau n^{2}\simeq 1$. This means that the problem Hamiltonian is of the
order of the level separation of the kinetic term only for sufficiently
small times. Since the eigenstates of the driving Hamiltonian are extended
(delocalized) in $m$, this implies the following qualitative picture of
adiabatic evolution of the ground energy level with $\tau $ . At
sufficiently small times $\tau n^{2}\ll 1$, the term due to $H_{P}$ can be
considered as a perturbation to the driver term and the energy levels are
not strongly distorted. The eigenstates are delocalized in the $m$ space. As
$\tau $ increases, the ground state is affected by the perturbation and the
ground state energy increases.

\begin{figure}
\begin{center}
\includegraphics[bb=-50 10 341 178, width=3.7in]{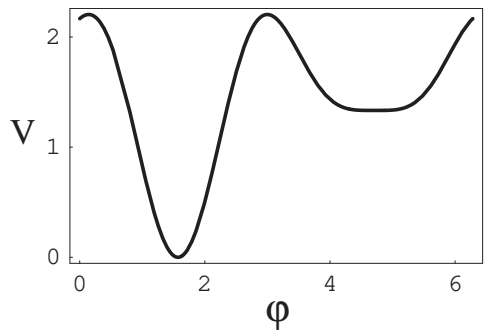}
\caption{\label{fig:V} Multistable effective potential $V$ {\it
vs} polar angle $\varphi$ for $q=3$.}
\end{center}
\end{figure}
\subsection{Minimum gap analysis}
If the ground state is not degenerate (this is the case when the total
number of qubits $n=2s$ is even), the gap may have a non-monotonic behavior
in $\tau $ in the range $\tau n^{2}\simeq 1$. Qualitatively, this can be
described as follows. For sufficiently small times $\tau n^{2}\ll 1$, the
ground state energy is increasing in $\tau $\ due to the diagonal matrix
elements of the problem Hamiltonian until compensated by the level repulsion
from the first excited state. After this, the ground state is pushed down
and gradually approaches the ground state of $H_{P}$ (which is $E_{g}=0$ in
the present case). Since the upper levels are not as strongly affected by
the perturbation as the ground state and since the number of upper levels is
very large, ${\cal O}(n)$, the system of upper levels behaves as a rigid one. For
this reason the strong interaction between the ground and first excited
states occurs in the range of energies $\Delta E$ corresponding to the the level separation in
 the driver Hamiltonian and is linear in $n$. Therefore
\begin{equation}
\Delta E ={\cal O}\left( n^{-2}\Delta E_{P}\right),  \label{gap_1}
\end{equation}
\noindent
where $\Delta E_{P}$ denotes the separation between the two lowest
eigenvalues of $H_P$. Clearly, the energy scale of the
minimal gap is polynomial in the number of qubits $n$ provided that
$\Delta E_{P}$ is polynomial.

As we have discussed above, in case of an  odd number of qubits,
all initial eigenstates of $H_{D}$ have a twofold degeneracy
(Kramers' degeneracy). In this case, the degeneracy is removed by
the effective potential (Zeeman splitting) and there is no
avoided-crossing  between the ground state and the first
excited states of the total Hamiltonian $H(\tau)$. One can easily show that
in this case the level separation $E_{1}(\tau)-E_{0}(\tau)$ 
grows  monotonically in $\tau $ \cite{LANDAU1} (it is linear in $\eta$
for $\eta\ll 1$).

Apart from the features occurring in the range of small $\tau \simeq n^{-2}$%
, the evolution of the ground state on the large time scale $\tau \simeq 1$
is the same for both even and odd number of qubits. Most importantly, the
estimate (\ref{gap_1}) holds globally in both cases. As we will see below,
this is confirmed by numerical simulations.

\section{\label{sec:gap1}Minimal Gap Estimate}

As we discussed above, the ground state energy may have a non-monotonic time
dependence on $\tau $ when the total number of qubits is even and the ground
state of $H_{D}$ is not degenerate. Because the eigenvalues of the driver Hamiltonian, $n\, k^2$, grow rapidly with the quantum number $k$ it is possible to
obtain a minimum gap estimate  analyzing how  
$H_{P}$ affects the two lowest
eigenvalues of the driver.  Making use of (\ref{H_tot1}), we obtain the 
adiabatic gap as a
function of time in the range $\tau n^{2}\lesssim 1$
\begin{equation}
E_{1}(\tau)-E_{0}(\tau) \approx \sqrt{\left[ \left( 1-\tau \right)
n-\tau \left( H_{P}\right) _{e_{1},e_{2}}\right] ^{2}+8\tau ^{2}\left|
\left( H_{P}\right) _{g,e}\right| ^{2}},  \label{pert_1}
\end{equation}
Here  we
explicitely take into account that the first excited level of $H_D$
is twofold degenerate. Ssubscripts $g$, $e_{1,2}$ above denote the ground state of $H_D$ and the two
lowest exited states, respectively. Matrix elements of $H_P$ on these states
satisfy the following relations:
\begin{equation}
\left( H_{P}\right)_{g,g}=\left( H_{P}\right) _{e,e}, \qquad \left( H_{P}\right)
_{g,e_{1}}=\left( H_{P}\right)_{g,e_{2}}\equiv \left( H_{P}\right)_{g,e}.
\end{equation}
\noindent
From (\ref{pert_1}), we obtain an estimate for the time when the minimal gap
is achieved
\begin{equation}
\tau _{c}=\frac{n\left[ \left( H_{P}\right) _{e_{1},e_{2}}+n\right] }{\left|
\left( H_{P}\right) _{g,e}\right| ^{2}+\left[ \left( H_{P}\right)
_{e_{1},e_{2}}+n\right] ^{2}},
\end{equation}
\noindent
In the limit $n\gg 1$ we can write
\begin{equation}
\eta _{c}=\tau _{c}n^{2}\approx \frac{n^{3}\left( H_{P}\right) _{e_{1},e_{2}}%
}{\left| \left( H_{P}\right) _{g,e}\right| ^{2}+\left[ \left( H_{P}\right)
_{e_{1},e_{2}}\right] ^{2}}.
\end{equation}
\noindent
Since matrix elements of $H_{P}\varpropto n^{3}$, it can be easily verified that the scaled
quantity $\eta _{c}$ does not depend on $n$. The matrix elements can be
calculated either in the quasiclassical basis given by (\ref{Schr_qc_2}) or
exactly using (\ref{S_x_21}) (see Appendix). For the Hamiltonian (\ref
{S_x_21}), the quasiclassical matrix elements yield
\begin{equation}
\left[ \left( H_{P}\right) _{e_{1},e_{2}}\right] ^{2}=\frac{9}{1024}n^{6}.
\end{equation}
\noindent
(for $q=3$). Therefore, we have $\eta _{\min }=\frac{64}{9}\approx 7.1$. This is in
a qualitative agreement with $\eta _{\min }\approx 5.0$ obtained in our
numerical simulations. Substituting into (\ref{pert_1}), we obtain the
estimate for the minimal gap
\begin{equation}
\Delta E_{\min }=\Delta E\left( \tau _{c}\right) =n\left( \frac{2}{3}\right)
^{1/2}.  \label{gap_pert_1}
\end{equation}
\noindent
The corresponding value of the slope $\Delta E_{\min }/n=\sqrt{\frac{2}{3}}\approx
0.82$ is again in a qualitative agreement with the value $0.86$ obtained in
the numerical simulations (Fig.5).

\begin{figure}[ht]\hspace{-0.33in}
\includegraphics[bb=-50 10 341 178,width=3.7in]{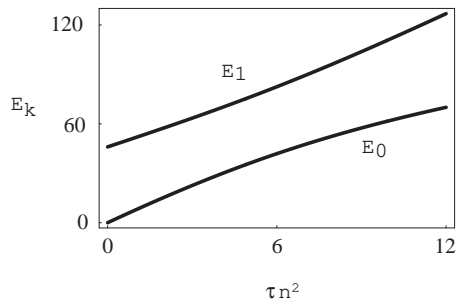}
\caption{\label{fig:Ek} Two lowest eigenvalues of $H(\tau)$
(\ref{S_x_21}): $E_{0,1}$  {\it vs} $\tau\,n^2$ in the vicinity of
avoided crossing for $n=46$ and $q=3$.}
\end{figure}

\begin{figure}[ht]
\includegraphics[bb=-50 10 341 178, width=3.7in]{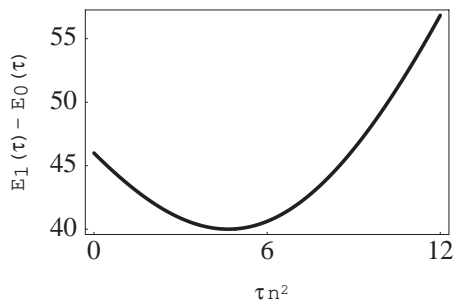}
\caption{\label{fig:Ek1} Difference between the two lowest
eigenvalues of $H(\tau)$  (\ref{S_x_21}): $E_{1}-E_{0}$   {\it vs}
$\tau\,n^2$ $n=46$ and $q=3$.}
\end{figure}

\subsection{Numerical analysis}
We also performed numerical simulations of the adiabatic spectrum
with Hamiltonian  (\ref{S_x_21}). In Fig.~\ref{fig:Ek}, we
plot the ground state and the first excited state energies as
functions of the dimensionless ''time'' $\eta =\tau n^{2}$ for
even values of $n$. According to the above discussion, in this case
the ground state is not degenerate and the energy level differnce 
 exhibits a
non-monotonic behavior (cf. Fig.~\ref{fig:Ek1}) leading to the polynomial minimal gap
$\Delta E_{\min}\sim n$ at $\eta \sim 1$. 
\begin{figure}[ht]
\includegraphics[ width=4in]{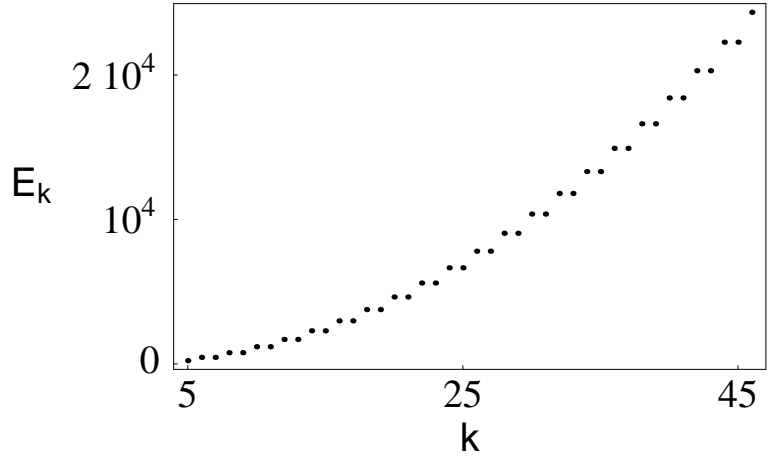}
\caption{\label{fig:Ek2} Eigenvalues of the
Hamiltonian $H(\tau)$ (\ref{S_x_21}) at the avoided-crossing,
$\tau=\tau_c$; $n=46$, $q=3$.}
\end{figure}
\noindent
In Fig.~\ref{fig:Ek2} we plot an eigenvalue spectrum of the
Hamiltonian $H(\tau)$ at the avoided crossing point,
$\tau=\tau_c$ for an even value of $n$.  For 
 $k\gg 1$ $E_{k}={\cal O}(n^3\,k^2)$ which corresponds to the scaling analysis in
Eqs.~(\ref{scale_1}),(\ref{scale_2}). Zeeman splitting
of a  doublet of the two lowest exited energy levels 
in Fig.~\ref{fig:Ek2}  is of the order of $n^3\tau_c={\cal O}(n)$.

In Fig.3 we plot the dependence of  minimal gap $\Delta E_{\min}$
{\it vs} $n$ for even values of $n$ .  The insert to this plot
corresponds to the dependence of the avoided crossing point $\tau_{c}$
{\it vs} $n$.

Using the simple  quasi-classical picture presented above it is
not difficult to compute the minimum gap in the case of the driver
Hamiltonian (\ref{HDfarhi}) in terms of the appropriate tunneling
exponent and recover the answer given in \cite{annealing}.

\begin{figure}[ht]\hspace{-0.33in}
\includegraphics[width=3.7in]{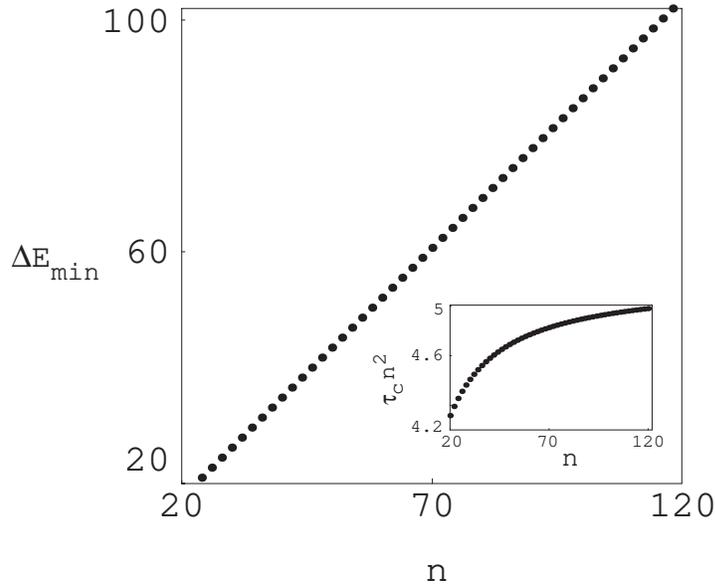}
\caption{\label{fig:gap} Minimum gap {\it vs} $n$ for even $n$.
The slope of the graph is 0.86 corresponding approximately to the analytical
estimate for $\Delta E_{\min}=(2/3)^{1/2}$ (\ref{gap_pert_1}). 
{\it Insert:} Scaled position of the
avoided crossing point $\tau_{c}n^2$ {\it vs} $n$. The assymptotic result 
is $\tau_c\,n^2\approx 4.98$ at $n\gg 1$.}
\end{figure}

\section{\label{sec:conclusion} Conclusion}

We show that macroscopic tunneling in QAA with the symmetrized cost
function can be totally suppressed using a driver Hamiltonian with the
ground state extended in the space of a total spin projection onto the
direction of computational basis. This leads to a polynomial time
complexity of QAA.  We give a simple form of a driver Hamiltonian
(\ref{driverSx2}) that has the aforementioned property and therefore
makes the algorithm polynomial.

We developed a simple intuitive picture of this phenomenon based on
WKB approximation for large spins. It follows from the analysis that
the results will hold for {\em any} form of $f(w)$ that is
sufficiently smooth on the scale $w\sim 1$.  We conjecture that this
phenomenon holds even if the cost function $E_{\bf z}$ is symmetrical
with respect to the permutation of bits but is not expressed only
throught the Hamming weight of a string.

We also argue using a picture of QAA as a quantum local search
\cite{Vazirani02,ST} that suppression of tunneling barriers with the
operator $S_{x}^{2}$ changes the global neighborhood properties for
QAA in a very profound way that can have an effect on the algorithm
complexity for a larger class of cost functions, $E_{\bf z}=f(w_{\bf
  z})+\Delta f_{\bf z}$ where $ \Delta f_{\bf z}$ breaks the symmetry
between the bits. In particular, it can be large for those states
$|{\bf z^{\prime}}\rangle$ that have exponentially small overlap
$|\langle {\bf z^{\prime}}|\Psi_{0}(\tau)\rangle|$ with adiabatic
ground states at all times (cf.  Sec.~7.3 of \cite{Vazirani02}).

A possible generalization of the above analysis is related to random
optimization problems with frustration, such as NP-hard problems and
corresponding spin glass models.  Exponential complexity of quantum
adiabatic evolution algorithms for these problems is not necessarily
related to tunneling but rather to the quantum diffusion phenomenon
associated with the rapid falloff of correlations in the bit-structure
with growing size of neighborhood around a given string \cite{ST}.
The role of the tunneling and collective phenomena involving
the low cost configurations in the performance of the quantum
adiabatic evolution algorithms in random NP-hard is yet to be analyzed.

\section{Acknowledgments}
We wish to thank Edward Farhi, Samuel Gutmann, Andrew Childs ( MIT),
and Umesh Vazirani ( UC Berkley) for stimulating discussions.  We also wish
to thank Alex Burin (Northwestern University) for turning our attention to
Refs.~\cite{GARG1,GARG2}. This research was supported by NASA
Intelligent Systems Revolutionary Computing Algorithms program
(project No:749-40).

\section{\label{sec:note} Note Added}
Shortly after this work was completed, we learned about the work of
E.Farhi, J. Goldstone, S. Gutmann \cite{farhi_paths} wherein another
approach was suggested to achieve polynomial complexity of the quantum
adiabatic evolution algorithm for the same optimization problem with
the symmetrized cost function (\ref{f0}). The algorithm proposed in
Ref.~\cite{farhi_paths} is quite different from our algorithm. We
believe that the proposal in \cite{farhi_paths} for random generation
of interpolating \lq\lq paths" $H(t)$ in different trials of the
algorithm is a very promising tool. It can perhaps be modified, by
including intermediate measurents \cite{farhimeas}, to become an
efficient adaptive algorithm for random NP-hard optimization problems.

However we believe the particular method of achieving polynomial
complexity presented in \cite{farhi_paths} is less robust than ours in
the problems with $n$ qubits where the dynamics of the total spin
$n/2$ is a key. The difference between the two algorithms is that in
our approach a simple universal form of the driver Hamiltonian {\em
  guarantees} that the minimum gap scales polynomially in the problem
size for a broad class of symmetrized cost functions $f(w)$. Also, by
construction our method does not require a specific knowledge of the
solution, or even a specific form of the cost function $f(w)$.

We also believe that simple quasi-classical approach presented above
enables one to study analytically a \lq\lq volume" in the space of
possible interpolating paths \cite{farhi_paths} that find a solution
in polynomial time. Detailed study of this issue will be presented
elsewhere.

\section{Appendix: Minimum gap estimate using matrix representation}

The eigenstates and eigenvectors of the Hamiltonian (\ref{S_x_21}) can be
obtained solving from the matrix representation. Choosing $x$ as a
quantization axis, we have $S_{z}=\left( -i\right) \left( S_{+}-S_{-}\right)
$ and
\begin{eqnarray*}
\left( S_{x}^{2}\right) _{m,m} &=&m^{2},\quad m=-l,-l+1,\ldots,l, \\
\left( S_{+}\right) _{m,m-1} &=&\left( S_{-}\right) _{m-1,m}=\frac{1}{2}%
\sqrt{\left[ l\left( l+1\right) -m\left( m-1\right) \right] }.
\end{eqnarray*}
\noindent
The full Hamiltonian (\ref{S_x_21}) is given by

\begin{eqnarray}
H\left( \tau \right) &=&\left( 1-\tau \right) nS_{x}^{2}+\tau \left( \frac{n%
}{2}\right) ^{3}h\left( S_{z}\right) ,  \nonumber \\
h\left( S_{z}\right) &=&\frac{1}{2}\left[ \left( q+\frac{4}{3}\right) +\frac{%
\left( 2-q\right) }{l}S_{z}-\frac{q}{l^{2}}S_{z}^{2}+\frac{\left(
q-2/3\right) }{l^{3}}S_{z}^{3}\right] ,  \nonumber
\end{eqnarray}
\noindent
and has the following matrix elements

\begin{eqnarray}
\left( H\right) _{m,m} &=&\left( 1-\tau \right) n\ m^{2}+\tau \left( \frac{n
}{2}\right) ^{3}\frac{1}{2}\left( q+\frac{4}{3}-q\Lambda _{0}\right) ,
\nonumber \\
\left( H\right) _{m,m-1} &=&\frac{1}{2}\tau \left( \frac{n}{2}\right) ^{3}
\left[ \left( 2-q\right) \Lambda _{1}+\left( q-2/3\right) \Lambda _{2}\right]
,  \label{matr_1} \\
\left( H\right) _{m,m-2} &=&\frac{1}{2}\tau \left( \frac{n}{2}\right)
^{3}\left( -q\right) \Lambda _{3},  \nonumber \\
\left( H\right) _{m,m-3} &=&\frac{1}{2}\tau \left( \frac{n}{2}\right)
^{3}\left( q-2/3\right) \Lambda _{4},  \nonumber
\end{eqnarray}
\noindent
with

\begin{eqnarray*}
\Lambda _{0}\left( l,m\right) &=&\frac{1}{2l^{2}}\left[ l\left( l+1\right)
-m^{2}\right] , \\
\Lambda _{1}\left( l,m\right) &=&\frac{\left( -i\right) }{2l}\sqrt{l\left(
l+1\right) -m\left( m-1\right) }, \\
\Lambda _{2}\left( l,m\right) &=&\frac{3i}{4l^{3}}\sqrt{l\left( l+1\right)
-m\left( m-1\right) }\left[ l\left( l+1\right) -m\left( m-1\right) -1\right]
, \\
\Lambda _{3}\left( l,m\right) &=&-\frac{1}{4l^{2}}\sqrt{\left[ l\left(
l+1\right) -m\left( m-1\right) \right] \left[ l\left( l+1\right) -\left(
m-1\right) \left( m-2\right) \right] }, \\
\Lambda _{4}\left( l,m\right) &=&-\frac{i}{2l}\sqrt{\left[ l\left(
l+1\right) -\left( m-2\right) \left( m-3\right) \right] }\Lambda _{3}\left(
l,m\right) .
\end{eqnarray*}
\noindent
One should note that the coefficients $\left\{ \Lambda _{k}\right\} $ do not
scale with $l=2n$, i.e. $\Lambda _{k}={\cal O}(1)$ for $n\gg1$. From (\ref
{matr_1}), it follows that the eigenvalue problem $H\left( \tau \right) \Psi
_{k}=E_{k}\left( \tau \right) \Psi _{k}$ is rescaled for $n\gg 11$ in terms of
dimensionless variables $\eta =\tau n^{2}$ and $\widetilde{E}_{k}\left( \tau
\right) =E_{k}\left( \tau \right) /n$ as

\begin{equation}
\widetilde{H}\left( \tau \right) \Psi _{k}=\widetilde{E}_{k}\left( \tau
\right) \Psi _{k},  \label{scale_1}
\end{equation}
\noindent
with
\begin{eqnarray}
\left( \widetilde{H}\right) _{m,m} &=&m^{2}+\frac{\eta }{16}\left( q+\frac{4
}{3}-q\Lambda _{0}\right) ,  \nonumber \\
\left( \widetilde{H}\right) _{m,m-1} &=&\frac{\eta }{16}\left[ \left(
2-q\right) \Lambda _{1}+\left( q-2/3\right) \Lambda _{2}\right] ,
\label{scale_2} \\
\left( \widetilde{H}\right) _{m,m-2} &=&\frac{\eta }{16}\left( -q\right)
\Lambda _{3},  \nonumber \\
\left( \widetilde{H}\right) _{m,m-3} &=&\frac{\eta }{16}\left( q-2/3\right)
\Lambda _{4},  \nonumber
\end{eqnarray}
\noindent
implying that matrix elements of $\widetilde{H}$ do not scale with $n$
and neither do the eigenvalues of $\widetilde{H}$.  It is
strighforward to obtain an estimate for the minimum gap
from the above equation using a Brillouin-Wigner perturbation theory in
parameter $\eta$. If one uses a 3-level truncation scheme,
 ($m=0,\pm 1$), and sets $q=3$ 
then a quasi-classical estimate (\ref{gap_pert_1}) 
can be recovered to the leading order in $n$.


\begin{thebibliography}{99}
\bibitem{footnote1}
When this work was completed we read the paper \cite{farhi_paths}
in which a quite different proposal for the polynomial time QAA
 was suggested for the same type of optimization
problems. We provide a brief discussion and comparision of the
approaches in Sec.~\ref{sec:note}.
\end{thebibliography}
\end{document}